\documentclass[10pt,prl,aps,showpacs,twocolumn]{revtex4-1}
\usepackage{epsfig}

\begin{document}

\title{Quantum Light from a Whispering Gallery Mode Disk Resonator}

\author{J. U. F{\"u}rst$^{1,2}$}
\author{D. V. Strekalov$^{1,3}$}
\author{D. Elser$^{1,2}$}
\author{A. Aiello$^{1,2}$}
\author{U. L. Andersen$^{1,4}$}
\author{Ch. Marquardt$^{1,2}$}
\author{G. Leuchs$^{1,2}$}

\affiliation{$^1$Max Planck Institute for the Science of Light, Erlangen, Germany\\ $^2$Department of Physics, University of Erlangen-Nuremberg, Germany\\
$^3$Jet Propulsion Laboratory, California Institute of
Technology, Pasadena, California, USA\\
$^4$Department of Physics, Technical University of Denmark, Kongens Lyngby, Denmark}

\date{\today}

\begin{abstract}
Optical parametric downconversion has proven to be a valuable source of nonclassical light. The process is inherently able to produce twin beam correlations along with individual intensity squeezing of either parametric beam, when pumped far above threshold. Here, we present for the first time the direct observation of intensity squeezing of -1.2dB of each of the individual parametric beams in parametric downconversion by use of a high quality whispering gallery mode disk resonator. In addition we observed twin beam quantum correlations of -2.7dB with this cavity. Such resonators feature strong optical confinement, and offer tunable coupling to an external optical field. This work exemplifies the potential of crystalline whispering gallery mode resonators in quantum and nonlinear optics and in particular for the generation of quantum light. The simplicity of this device makes the application of quantum light in various fields highly feasible.
\end{abstract}

\maketitle


When going back and forth on a swing, the oscillatory movement of the legs is coupled to the oscillation of the swing \textendash$\,\,$a well-known parametric process. Similarly, optical subharmonic generation, also referred to as parametric downconversion (PDC), links an optical field to its subharmonic mediated by a dielectric medium. In this optical process, one pump photon (p) is converted to two subharmonic photons, called signal (s) and idler (i). The signal and idler are therefore strongly correlated in photon number. Experimentally, these two-mode correlations were proven to be quantum correlations \cite{fabre_twinbeam,Wu_PDCsqueeze}. Further improvement made this optical parametric process a state-of-the-art squeezing source \cite{Laurat05,mehmet,coelho09}, along with squeezing in optical fibers \cite{shelby,dong}.

About twenty years ago, a theoretical analysis additionally predicted the generation of entangled signal and idler, which are simultaneously individually squeezed in intensity, when pumping the process high above the optical parametric oscillations (OPO) threshold \cite{variance,drummond}. However, this could not be shown directly so far, as the above-threshold squeezing was obscured by classical noise of the pump laser source and by relaxation oscillations occuring in triply resonant cavities \cite{relaxosc,Porzio}. In our approach, we were able to overcome these obstacles by utilizing a compact and robust whispering gallery mode (WGM) resonator, which is a valuable tool in many areas in photonics \cite{ilchenko_basics,vahala1,ilchenko_app}.
 In these resonators, light is guided by continuous total internal reflection. This lifts the need for high reflectivity coatings, thereby limiting the quality of WGM resonators by internal material loss only \cite{Savchenkov042}. The evanescent field overlap with an external optical element allows for continuously variable coupling to the WGM resonator. So far, the extreme properties of WGM resonators could not be leveraged to generate quantum light from nonlinear effects of the resonator material, as e.g. thermorefractive noise played a detrimental role. By using the high second order nonlinearity in a crystalline WGM resonator, we succeeded to show quantum two-mode correlations in the downconverted beams. In addition, the extremely low threshold enabled us to directly show the squeezing of a single parametric beam in this experiment.



Cavity properties \textendash$\,\,$in particular of high quality cavities, as our WGM resonator \textendash$\,\,$significantly influence the quantum and classical properties of the optical parametric oscillations (OPO) \cite{Aiello}. In the following, we discuss theoretical implications of a triply resonant degenerate OPO pumped coherently above threshold (see \cite{variance}). The resonator is characterized by the total pump and parametric cavity linewidth $\gamma_p$ and $\gamma$, respectively. The total cavity linewidth $\gamma$ is equal to the sum of the internal loss rate $\alpha$ and the coupling rate $\gamma_0$ (we refer to energy loss rates, because they are equal to the cavity linewidth).

The OPO threshold pump power $P_{th}$, for instance, depends cubically on the total cavity linewidth of the three fields involved, as $P_{th}\propto\gamma_p \,\gamma^2$. We expect an extremely low threshold, as WGM resonators are limited by absorption loss only. Hence, only a low optical pump power $P$ is required, where shot noise limited lasers are easily available. Moreover, we are able to continuously tune the threshold by varying the coupling rate.

The twin-beam intensity correlations between signal and idler are affected by the presence of a cavity as well. These correlations are proportional to the variance of the difference $\mathcal{V}_-$ of the two beams field amplitude quadratures $\hat X_s$ and $\hat X_i$. The variance of the quadrature sum $\mathcal{V}_+$ yields the total noise of the parametric beams. Theory for these variances normalized to the shot noise limit (SNL) predicts \cite{variance}
\begin{equation}
\mathcal{V}_-=1- \frac{\gamma_0}{\gamma}\frac{\eta}{1+\Omega^2},\,
\mathcal{V}_+=1+ \frac{\gamma_0}{\gamma}\frac{\eta}{(\sigma-1)^2+\Omega^2}.
\label{equ:twin}
\end{equation}
 Here, $\Omega=\nu_{det}/\gamma$ is the measurement sideband frequency $\nu_{det}$ normalized to the parametric cavity bandwidth. The pump parameter is $\sigma = \sqrt{P/P_{th}}$ and the detection efficiency $\eta$ includes all optical losses in the signal (or idler) channel. 
The difference signal $\mathcal{V}_-$ is independent of $\sigma$ and shows twin beam quantum correlations, as $\mathcal{V}_-<1$. The sum signal $\mathcal{V}_+$ on the contrary depends on the pump parameter. It shows strong excess noise ($\mathcal{V}_+\gg1$) near the threshold and drops to  the SNL for high pump powers. Both variances depend on the ratio of parametric coupling rates to total losses. With a high quality WGM OPO, we are able to vary this ratio between coupling and internal losses for optimizing the correlations, while maintaining a reasonably low threshold. Thus, shot noise limited or even non-classical pump fields are accessible. 

The individual intensity fluctuations of the signal and equivalently the idler field depend strongly on the cavity as well. They are proportional to the normalized amplitude quadrature variance \cite{variance}
\begin{equation}
\mathcal{V}(\hat X_{s,i})=1-\frac{\eta}{2} \frac{\gamma_0}{\gamma}\frac{\sigma(\sigma-2)}{(1+\Omega^2)(\Omega^2+(\sigma-1)^2)}.
\label{equ:single}
\end{equation}
Contrary to the twin beam correlations $\mathcal{V}_-$, the single beam intensity variance $\mathcal{V}(\hat X_{s,i})$ strongly depends on the pump parameter and drops from excess noise near the threshold, to an ideal squeezing limit of 0.5 (-3 dB) far above the threshold. The SNL is reached at four times the threshold. Analogous to the twin beam correlations, the individual squeezing depends on the ratio of coupling rates to the total loss.  To reach a high pump parameter however, one needs to keep the threshold low, which implies a high quality resonator.  Additionally, one has to assure a coherent pump field for the observation of squeezing, easily accessible at low pump powers. In this respect, the quality of a WGM resonator gives an additional significant advantage.


Triply resonant OPOs are also known to undergo relaxation oscillations \cite{Lee,Porzio}. These induce classical fluctuations in the AC power spectrum of each parametric field at a specific frequency, and can disturb the measurement of squeezing. The relaxation oscillation frequency  $\nu_R$ normalized to the parametric linewidth $\gamma$ is given by \cite{Lee}
\begin{equation}
\nu_N=\frac{\nu_R}{\gamma}=
\sqrt{\frac{\gamma_p}{2\gamma}}\cdot\sqrt{\sigma-\left( 1+ \frac{\gamma_p}{4\gamma}\right)}.
\label{eq:relax}
\end{equation}
The relaxation oscillations have a threshold at $\sigma=1+\gamma_p/(4\gamma)$ and depend on the pump parameter $\sigma$. One possible approach to circumventing the relaxation oscillations is to make its threshold very high by having a large ratio of $\gamma_p/\gamma$, i. e. working with a cavity only resonant for the parametric fields. However, we cannot tune both loss rates independently, with the present coupling technique.  Alternatively, one can shift the relaxation oscillation frequency out of the cavity bandwidth ($\nu_N\gg 1$), by increasing the pump power. For the cavity used in this experiment, we estimated the normalized relaxation oscillations versus $\sigma$.
 The relaxation oscillations only lie within our cavity bandwidth for pump parameters $\sigma$ between approximately $2.5$ and $2.8$. Thus we are able to easily shift these oscillations out of the cavity bandwidth. 


For our experiment we used a WGM disk resonator, which posseses axial symmetry. To assure phase matching for PDC all along the circumference of the resonator the crystal should be uniaxial, with the optical axis along the symmetry axis of the cavity. Hence, our WGM disk is made from a 5\% MgO-doped z-cut Lithium Niobate wafer. Natural Type I phase matching in the crystal can be achieved by tuning the refractive indices of the disk via temperature and the electro-optical effect, as discussed in \cite{furst10}. The disk resonator radius is 1.9 mm, the rim radius is  0.25 mm and its height is 0.5 mm. The surface of the disk is polished to provide a flat surface for total internal reflection. 
\begin{figure}[t]
	\centering
	\includegraphics[width=6cm]{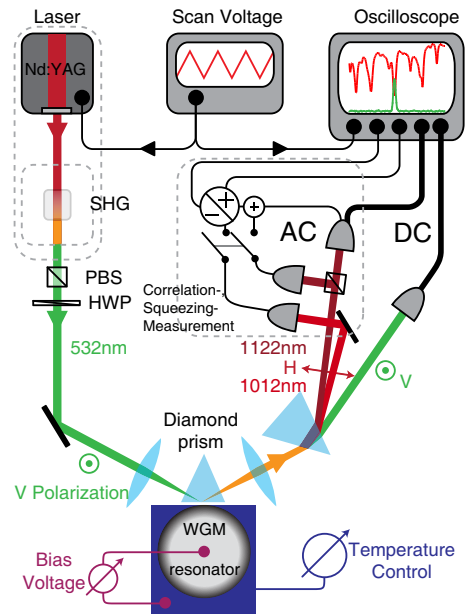}
	\caption{Schematic of the parametric downconversion setup.}
	\label{fig:setup}\vspace*{-0.2in}
\end{figure}
A sketch of our experimental setup is shown in Fig.~\ref{fig:setup}. We drive our triply resonant WGM optical parametric oscillation (OPO) above threshold with a continuous wave frequency doubled Nd:YAG laser at 532 nm. As our WGM OPO shows very good stability, no active locking is required. A movable diamond prism (antireflection coated for 532 nm \& 1064 nm) is placed in the vicinity of the WGM resonator, overlaping with its evanescent field. This allows for continuously variable coupling of the pump field to the resonator. The outcoupling is provided by the same prism. The outcoupled pump and strongly non-degenerate signal and idler fields are separated with a dispersion prism and directed to the detection setup, carefully assuring minimal losses for the PDC light. Continuously scanning the pump frequency, we observe the WGM spectrum and selected the pump mode most efficient for PDC. We applied a phase matching temperature near $94^{\circ}$C. The minimal threshold power for PDC was measured to be $6.7\,\mu$W, which is \textendash$\,\,$as expected \textendash$\,\,$extremely low, compared to state of the art experiments with $300\,\mu$W \cite{threshold} and gives us the necessary flexibility for the quantum measurements. The classical properties of our OPO are discussed in \cite{furstOPO}. 

For coupling the pump field critically to the resonator ($\gamma_{p,0}/\gamma_{p}=0.5$), the total cavity bandwidth of the pump field is observed to be $\gamma_p=30$ MHz  and its coupling rate is $\gamma_{p,0}=15$ MHz. Moreover, we found $\gamma_0/\gamma =0.22$ for the parametric light, using the OPO output power dependence on the pump parameter (see \cite{furstOPO}). This indicates, that signal and idler couple weaker to the resonator than the pump field. This is surprising, as the evanescent coupling is stronger for a fixed spacing, the longer the wavelength. However, we assume that the parametric WGMs excited in our OPO are non-equatorial, and deeper lying modes than the pump mode. By adjusting the phase matching conditions to excite parametric equatorial modes, we would expect stronger outcoupling, hence more squeezing.


For the quantum measurements, we use two photo detectors providing a good signal-to-noise ratio at low light powers (optical noise equivalent power is less than 1.5 $\mu$W) at a measurement frequency of $3.2$ MHz (resulting in a normalized cavity frequency of $\Omega = 0.6$). The DC part of the photo current is used as a monitor and for shot noise calibration. We simultaneously monitor the sum and the difference AC fluctuations of these two detectors with two spectrum analyzers configured for zero span around $3.2$ MHz, a resolution bandwidth of $300$ kHz, and a video bandwidth of $10$ kHz. These settings result in a time resolution of around 0.1 ms, while the pump laser sweep time is 50 ms. We postprocessed the spectrum analyzer data with a running average of $50$. The detection setup is linear in AC and DC over more than $17$ dB. The common mode rejection ratio between sum and difference channel is at least $20$ dB (i.e., the detectors are balanced to within 1\%). All squeezing measurements performed were verified by the linear behavior of attenuation measurements.

\begin{figure}
	\centering
	\includegraphics[width=7cm]{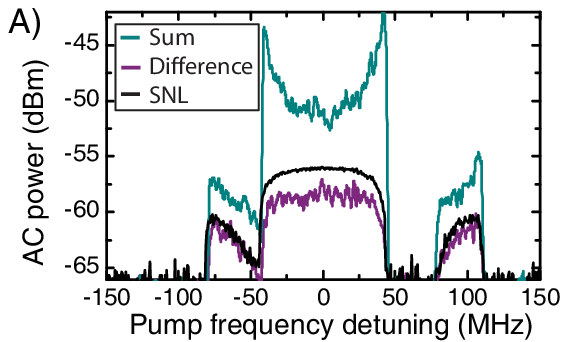}
	
	\vspace*{0.2in}
	
	\includegraphics[width=7cm]{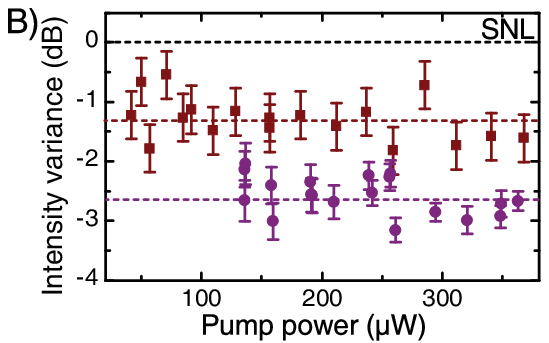}
	\caption{Two-mode squeezing in the twin beam measurement. {\bf A)} Variances of the signal and idler photo currents sum and difference compared to the SNL, vs. laser frequency detuning from the pump WGM center. {\bf B)} Twin beam squeezing vs. pump power for stronger coupling (more squeezing) 
and weaker coupling (less squeezing). 
Error bars are determined from the raw data, as seen in A). 
Horizontal lines indicate the SNL and the mean squeezing values, for weak coupling $-1.3$ dB, for strong coupling $-2.7$ dB. }
	\label{fig:3}	
\end{figure}

The PDC process generates photons by pairs, leading to quantum correlations in the intensity of signal and idler \cite{fabre_twinbeam}. These two-mode correlations can be observed by focusing signal and idler on two detectors and analyzing the fluctuations of the balanced sum and difference photo currents. When the difference fluctuations are below the shot noise level, one verifies that the twin beams are quantum correlated. The quantum efficiency of the measurement is $\eta=(87\pm 4)\%$. For this twin-beam measurement we set the threshold of our resonator slightly below the fixed pump power by adjusting the coupling. In Fig.~\ref{fig:3}~A) we present a sweep of the pump frequency through the pump WGM resonance. We see three PDC channels, one at the center of the pump resonance and two on the wings. The difference signal in Fig.~\ref{fig:3}~A) is reliably below the SNL for the central channel, where we observe $(-2.7\pm 0.4)$ dB of two-mode squeezing (corrected for electronic noise). The sum signal greatly exceeds the SNL in this near-threshold measurement. In a separate measurement we observed that it approaches the SNL for very high pump powers, as predicted in Eq. \ref{equ:twin}.

We measured quantum correlations for different pump powers in two coupling regimes (see Fig.~\ref{fig:3}~B). The theoretical estimate of -0.6 dB 
for a critically coupled pump, where our signal and idler modes are undcoupled ($\gamma_0/\gamma =0.22$), 
is close to the measured value for weak coupling. Stronger coupling leads to an increased output rate relative to the loss rate. This increases the threshold, but also the two-mode squeezing. This result is consistent with our theoretical estimate yielding -2.0 dB of quantum correlations for signal and idler WGMs critically coupled ($\gamma_0/\gamma=0.5$). As stated in Eq. \ref{equ:twin}, the correlations are independent of the pump power in both regimes within the measurement uncertainty.

On the contrary, the intensity fluctuations of a \textit{single} parametric beam are expected to depend strongly on the pump power, as discussed. Investigations of squeezing high above threshold now is feasible. 
We directly measured the signal (or idler) intensity fluctuations with balanced self-homodyning, that delivers the SNL level in parallel.

\begin{figure}[b]
	\centering
	\includegraphics[width=7cm]{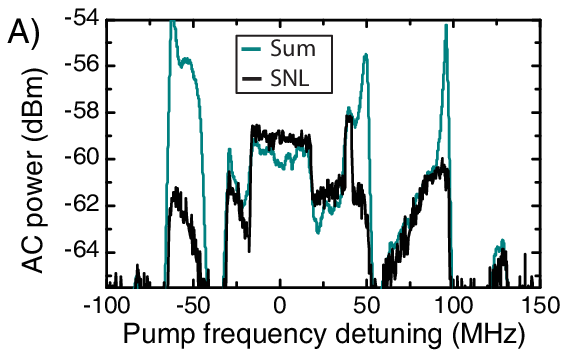}
	
	\vspace*{0.2in}
	
	\includegraphics[width=7cm]{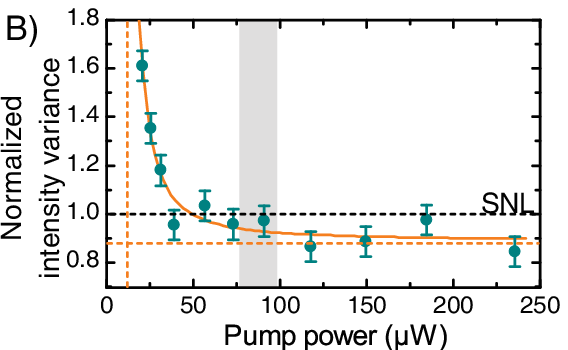}
	\caption{Single beam intensity noise measurement.  {\bf A)} Intensity noise compared to SNL vs. laser frequency detuning from the pump WGM center. {\bf B)} Intensity noise vs. pump power (measurement and theoretical fit). Error bars are estimated as the mean variance of two measurements at every point. The horizontal lines indicate the SNL (black) and the asymptotic limit of 0.9 for the normalized intensity variance (orange). The vertical line visualizes the OPO threshold of $12.3$ $\mu$W. The gray area marks the range of the pump power, where relaxation oscillations may be excited.
}
	\label{fig:4}
\end{figure}

In the single beam measurement the overall detection efficiency is $73\pm4\%$. We varied the coupling to facilitate a very low threshold. In Fig.~\ref{fig:4}~A) the pump frequency is swept through the resonance while the intensity noise and the SNL is recorded. Again, multiple PDC channels are excited. The central channel shows $(-1.2\pm0.4)$ dB of squeezing (corrected for electronic noise) when pumped with approximately 400 $\mu$W. We however select a different conversion channel to measure the squeezing as a function of the pump power (see Fig.~\ref{fig:4}~B), as the available pump power was limited. Fig.~\ref{fig:4}~B) shows strong excess noise just above the threshold. As the pump power increases, the noise decreases and finally falls below the SNL, at which point the field becomes squeezed. A theoretical fit for the normalized intensity variance from Eq. \ref{equ:single}  
 yields a threshold power of 12.3 $\mu$W and converges to 0.9 shot noise units (-0.5 dB) for high pump powers. 
This is in good agreement with the theory for critically coupled parametric WGMs ($\gamma = 2\,\gamma_0$), where the squeezing limit is $V_{crit,s}(\hat X_1) = 0.84$ shot noise units (-0.7dB).
 The squeezing was not impeded by relaxation oscillations, as their frequency depends very strongly on the pump power, and lies within the cavity bandwidth only for a very narrow range of pump power (see Fig.~\ref{fig:4}~B
). This has enabled the first direct observation of sub-poissonian photon statistics in a single OPO beam above threshold.


This work illustrates the potential of our WGM resonator in the generation of quantum states of light and nonlinear optics in general. The signal and idler beams in our experiment are not only quantum correlated, but also individually intensity squeezed. In addition, in the far below-threshold regime, our WGM resonator is expected to be a highly efficient narrow bandwidth single photon source, compatible with atomic linewidths. High flexibility in wavelength selection for coupling the single photons to atoms could be achieved using quasi phase matching in periodically poled resonators \cite{ilchenko04SH,sasagawa}. As crystalline WGM resonators offer high mechanical Q-factors as well \cite{kippmech}, one could combine optomechanics with non-classical states of light. Furthermore, it is intruiging to explore the nonclassical properties of the light fields in second order nonlinear WGM resonators at high pump powers, where dynamic effects arise.

 

\begin{acknowledgments}
The authors would like to acknowledge funding from EU project COMPAS. D.V.S. and A. A. acknowledges funding from the Alexander von Humboldt foundation, and J.U.F. from IMPRS. J.U.F. and Ch. M. thank Alessandro Villar for helpful discussions. J.U.F thanks Thomas Bauer for discussions about relaxation oscillations.
\end{acknowledgments}

\end{document}